\documentclass[twocolumn,prl,floatfix,showpacs,superscriptaddress] {revtex4}
\usepackage{graphics}
\usepackage{epsfig}

\usepackage{color}

\usepackage{amssymb}
\usepackage{amsmath}
\usepackage{graphicx}
\usepackage{float}

\def \be{\begin{equation}}
\def \ee{\end{equation}}
\def \bew{\begin{widetext}\begin{equation}}
\def \eew{\end{equation}\end{widetext}}
\def \bmlett{\begin{mathletters}}
\def \emlett{\end{mathletters}}

\def \r{{\bf r}}

\newcommand{\alp}{\alpha}
\newcommand{\ZPF}{{\rm ZPF}}

\def\be{\begin{equation}}
\def\ee{\end{equation}}

\def\w01{\omega_{01}}
\def\r0{R_0}

\def\omegar{\omega_{\rm R}}

\def\omegal{\omega_{\rm L}}

\def\omegam{\omega_{\rm M}}

\def\nmO{\bar n_{\rm M}^O}
\def\nm{\bar n_{\rm M}}
\def\nmT{\bar n_{\rm M}^T}

\begin{document}

\title{Quantum Theory of Cavity-Assisted Sideband Cooling of Mechanical Motion}

\author{Florian Marquardt}
\affiliation{Department of Physics, Arnold-Sommerfeld-Center for Theoretical Physics, and Center for
NanoScience, Ludwig-Maximilians-Universit\"at M\"unchen,  Theresienstr. 37, 80333 Munich, Germany}
\author{Joe P. Chen}
\affiliation{Department of Physics, Yale University, PO Box 208120, New Haven, CT 06520-8120}
\author{A.A. Clerk}
\affiliation{Department of Physics, McGill University, 3600 rue University, Montreal, QC Canada H3A 2T8}

\author{S.M. Girvin}

\affiliation{Department of Physics, Yale University, PO Box 208120, New Haven, CT 06520-8120}

\begin{abstract}

We present a fully quantum theory describing the cooling of a cantilever coupled via radiation pressure to an
illuminated optical cavity. Applying the quantum noise approach to the fluctuations of the radiation pressure
force, we derive the opto-mechanical cooling rate and the minimum achievable phonon number. We find that
reaching the quantum limit of arbitrarily small phonon numbers requires going into the good cavity (resolved
phonon sideband) regime where the cavity linewidth is much smaller than the mechanical frequency and the
corresponding cavity detuning. This is in contrast to the common assumption that the mechanical frequency and
the cavity detuning should be comparable to the cavity damping.
\end{abstract}


\date{\today}
\maketitle

Recent years have seen significant advances in fabricating, measuring, and controlling mechanical systems on the
micro- and nanometer scale
\cite{2000_09_ParkMcEuen_C60,2001_08_ErbeBlick_ElectronShuttle,2003_01_HuangRoukes_GHzResonator,2003_07_KnobelCleland_NatureChargeSensing,2004_01_WeigKotthaus_SuspendedQDot,2006_08_Schwab_CPB_Molasses}.
These advances have opened the possibility of observing quantum effects in mechanical devices. A number of
experiments to explore the quantum regime of these systems have been proposed, including measuring their
coherence and  entangling them with other systems
\cite{1997_04_Mancini_MirrorCat,1999_05_Bose_Cat,2003_09_Marshall_QSuperposMirror,2002_04_Armour_ResonatorCat,2003_07_Braunstein_LightToResonator,2005_12_Pinard_EntanglingAndSelfCooling},
generating nonclassical states \cite{1992_Hilico_Squeezing,2000_05_Blencowe_SqueezingCantilever},
 and making quantum-limited measurements
 \cite{1992_BraginskyKhalili_QuantumMeasurement,2004_04_LaHayeSchwab_QLimitNanoResonator}.
Many of these proposals require the mechanical system to be cooled to its ground state, i.e. to temperatures
below $20$~mK even for $1$~GHz resonators. This is difficult or impossible using bulk refrigeration, and a more
promising means is to make use of non-equilibrium cooling techniques, analogous to the laser-cooling schemes
developed for trapped ions and neutral atoms
\cite{1979_10_WinelandItano_LaserCoolingPRA}.

Such schemes fall into two categories. In the first, cooling is achieved via an active feedback loop which is
used to cancel the cantilever's thermal motion
\cite{1953_Milatz_FeedbackCooling,1999_10_Cohadon_CoolingMirrorFeedback,2006_11_Bouwmeester_FeedbackCooling}.
This approach (including its quantum limits \cite{2001_10_Courty_QLimitFeedbackDamping}) has been considered
elsewhere. Here we will focus on the second category: passive, non-feedback-based cooling. In this approach the
cantilever displacement is coupled parametrically to a driven resonator (or two-level system). When the
frequency of the drive applied to the resonator is chosen appropriately, this parametric coupling ensures that
the cantilever is preferentially driven to lower energy states. The lowest temperature  to which this process
can cool the cantilever is determined by the resonator's quantum fluctuations (photon shot-noise), which drive
the cantilever randomly and compete with the cooling.

This balance has been considered theoretically for a few specific realizations of the cooling system: a
Cooper-pair box \cite{2004_03_MartinShnirman_ResonatorQubitSidebandCooling},  the superconducting
single-electron transistor
\cite{2005_11_Clerk_SSET_Cooling,2005_11_Blencowe_SET_NJP,2006_08_Schwab_CPB_Molasses},  quantum dots
\cite{2004_02_WilsonRae_LaserCoolinNanoResonator} and ions \cite{2004_12_TianZoller_IonNEMS}. In this paper we
present a quantum mechanical description of the self-cooling of a cantilever coupled to an optical cavity via
radiation pressure. This system is of particular interest because of its relative simplicity - the cantilever is
cooled by a single mode of the electromagnetic field. It is also of interest because this system has already
been realized experimentally, and recently produced very promising results on self-cooling
using both photothermal forces and radiation pressure forces
\cite{2004_12_ConstanzeKhaled_WithNote,2006_05_Harris_MicrocantileverMirror,%
2006_11_KarraiNewsViews,2006_05_AspelmeyerZeilinger_SelfCoolingMirror,2006_06_Paternostro_NJP_Spectrum,%
2006_07_Arcizet_CoolingMirror,2006_11_Kippenberg_RadPressureCooling,2006_12_NergisMavalvala_LIGO}. Cooling rates
and steady state temperatures for such a scheme (and of related systems like a driven LC circuit
\cite{2006_06_Wineland_CantiCoolingRFCircuit}) have been derived only (semi-)classically so far. Most notably,
Braginsky and Vyatchanin \cite{2002_02_BraginskyVyatchanin_Tranquilizer} considered the optomechanical damping
within a semiclassical theory. Here, we use the quantum noise approach to find simple and transparent, but fully
quantum-mechanical  expressions, starting directly from the spectrum of the force fluctuations. Our results are
valid both for the good-cavity regime (resolved mechanical sidebands) and the bad-cavity regime (unresolved
sidebands). Earlier $\mbox{(semi-)}$ classical calculations
\cite{2002_02_BraginskyVyatchanin_Tranquilizer,2004_12_ConstanzeKhaled_WithNote,2006_06_Wineland_CantiCoolingRFCircuit}
are reproduced in the appropriate limits. We show in particular that it should be possible to cool the cantilever to its
quantum mechanical ground state by choosing the cantilever resonance frequency much larger than the cavity
ringdown rate, a regime that has not been considered so far. Cooling the cantilever to its ground state will
enable the realization of many of the aforementioned tasks, as well as investigations into the optomechanical
instability
\cite{2005_06_Vahala_SelfOscillationsCavity,2005_07_VahalaTheoryPRL,2005_02_MarquardtHarrisGirvin_Cavity} in the
quantum regime. The results are complemented by an exact solution of the coupled equations of motion to account for
the ``strong cooling'' limit, where the cooling rate exceeds the cavity ringdown rate.  

Consider a mechanical degree of freedom $\hat{x}$ coupled parametrically with strength $A$ to the cavity
oscillator
\be
{\hat H} = \hbar(\omegar - A \hat x)\, [{\hat a}^\dagger \hat{a}-\langle{\hat a}^\dagger \hat{a}\rangle] + {\hat
H}_{M} + {\hat H}_{\rm drive}+ {\hat H}_\kappa + {\hat H}_\Gamma \label{eq:paramcoupling}
\ee
where $\omegar$ is the cavity resonance frequency at the equilibrium cantilever position $x=0$ in the presence
of the mean radiation pressure, ${\hat H}_M=\hbar \omega_M {\hat c}^\dagger {\hat c}$ is the mechanical
oscillator, ${\hat H}_{\rm drive}$ is the optical drive, ${\hat H}_\kappa$ represents the cavity damping, and
${\hat H}_\Gamma$ represents the mechanical damping. For a cavity of length $L$, the change in cavity length
with $x$ gives a radiation pressure coupling $A= +\omegar/L$.

Opening the port used to supply the classical drive to the cavity also admits vacuum noise.
%
Writing the cavity field in terms of its classical and quantum parts $\hat{a}(t) = e^{-i\omegal t} [\bar a +
\hat{d}(t)]$ yields the photon number autocorrelation function
\begin{eqnarray}
\tilde S_{nn}(t) &=& \langle \hat{a}^\dagger(t) \hat{a}(t) \hat{a}^\dagger(0) \hat{a}(0)\rangle - \langle
\hat{a}^\dagger(t) \hat{a}(t)\rangle^2\nonumber\\
 &=& \bar n e^{i\Delta t-\frac{\kappa}{2}|t|}
\label{eq:autocorrelation}
\end{eqnarray}
where $\Delta=\omegal-\omegar$ is the laser-cavity detuning and $\bar n$ is the mean photon number. Because the
number fluctuations are the result of interference between the classical drive amplitude and the vacuum
fluctuation amplitude, they decay at the amplitude decay rate $\kappa/2$, \emph{not} the energy decay rate
$\kappa$ as is sometimes naively assumed.


\begin{figure}[tbh]
\begin{center}
\includegraphics[width=0.48\columnwidth]{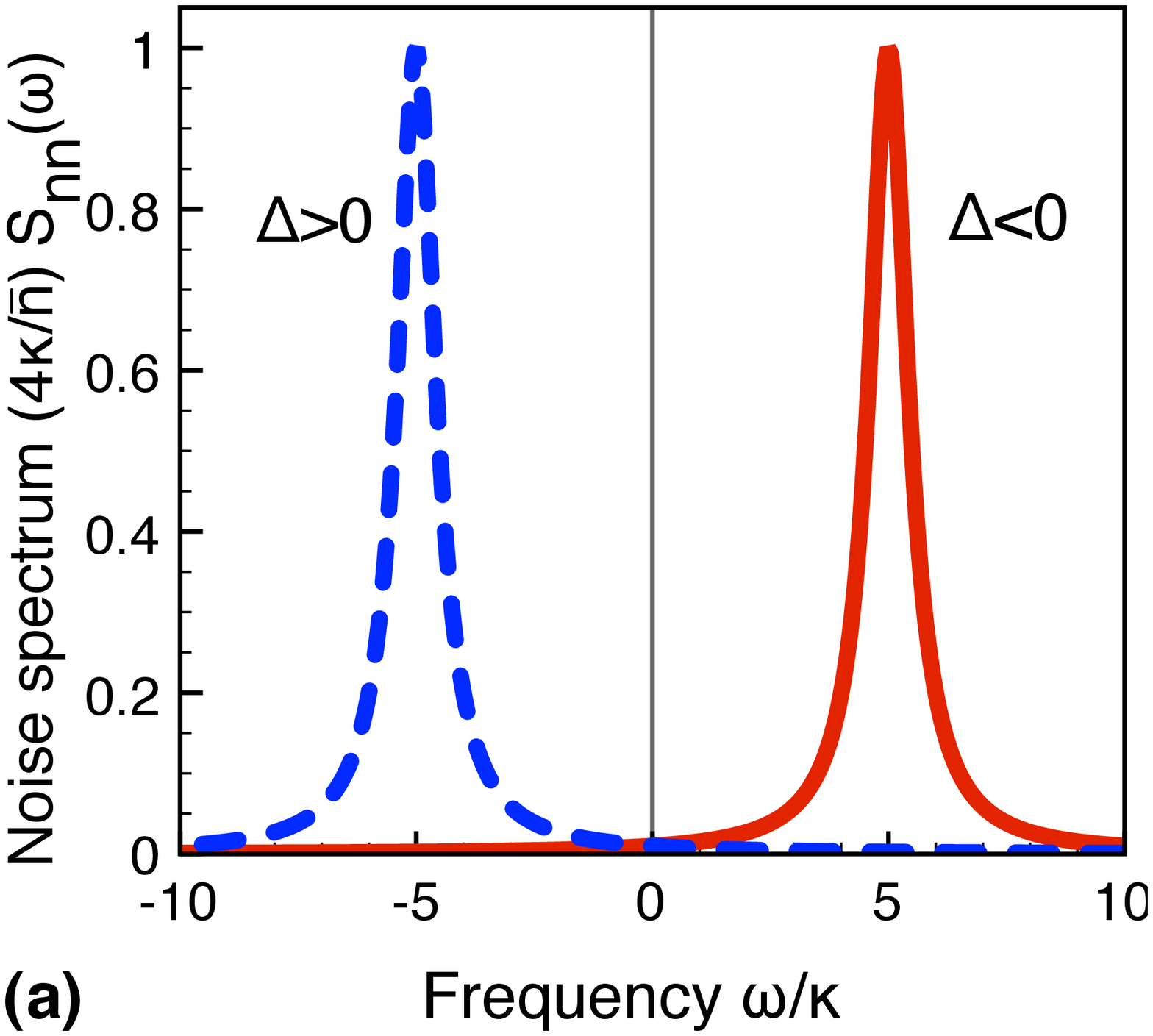}
\includegraphics[width=0.48\columnwidth]{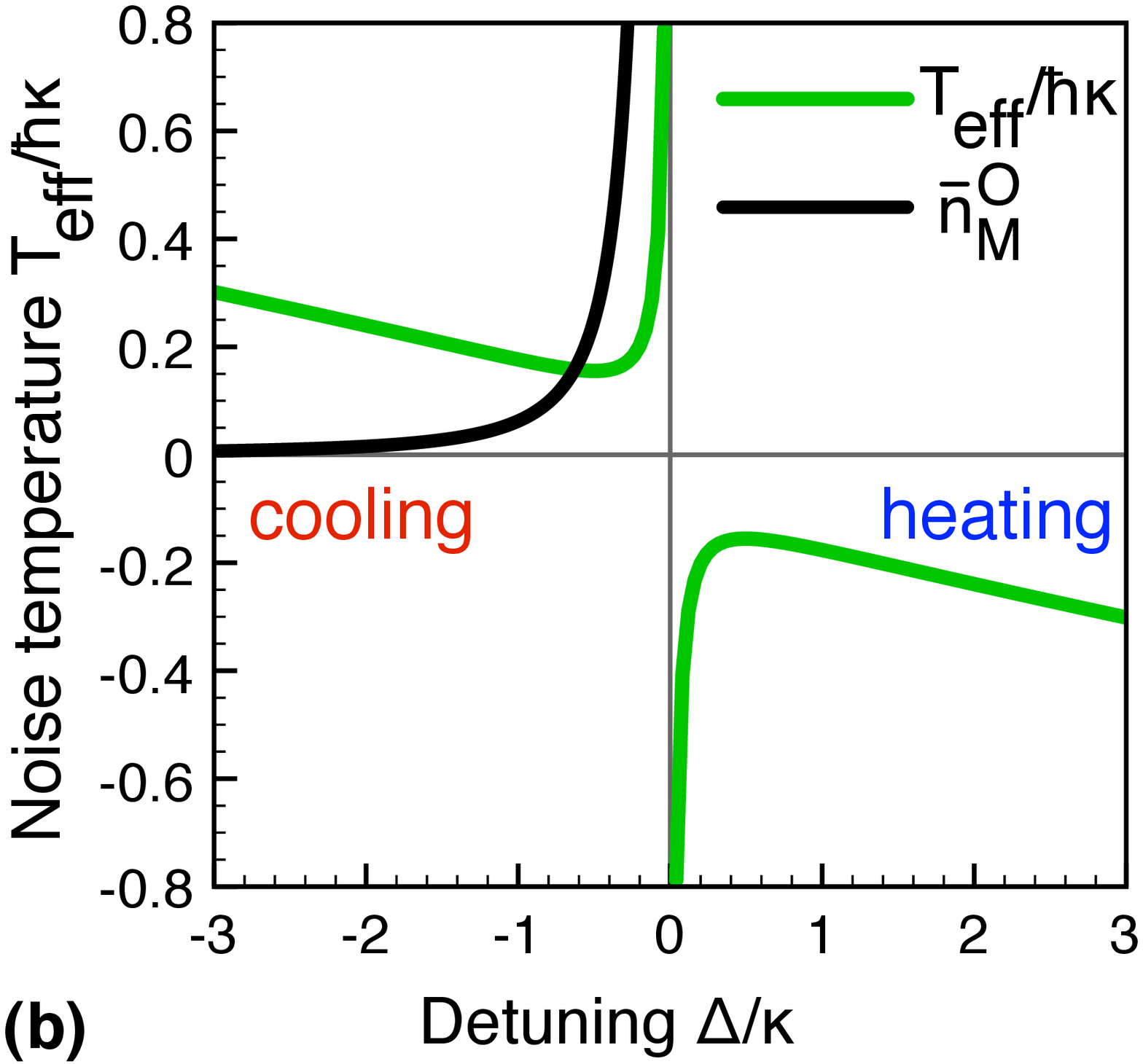}
\caption{(Color online) (a) Noise spectrum of the photon number in a driven cavity as a function of frequency when the cavity
drive frequency is detuned from the cavity resonance by $\Delta=+5\kappa$ (leading to heating: dashed line) and
$\Delta=-5\kappa$ (cooling: solid line). (b) Effective noise temperature $T_{\rm eff}$ as a function of the
detuning of the drive from the cavity resonance, see Eq. (\ref{nmO}), for a cantilever frequency matching the
detuning: $\omega_M=|\Delta|$. The noise temperature is positive on the left side of the resonance, where
optomechanical cooling produces a minimum reachable cantilever phonon number ${\bar n}^O_M$ (black line)
corresponding to $T_{\rm eff}$, see Eqs. (\ref{nmO},\ref{minimumPhononNumber}). } \label{fig:CavityTeff}
\end{center}
\end{figure}


The power spectrum of the noise is
\be
S_{nn}(\omega)=\int_{-\infty}^{+\infty} dt\, e^{i\omega t} \tilde S_{nn}(t)=\bar n \frac{\kappa}{(\omega+\Delta)^2 +
(\kappa/2)^2}. \label{eq:Snnomega}
\ee
The symmetric part of the noise spectrum was discussed in \cite{2002_02_BraginskyVyatchanin_Tranquilizer} to
estimate the additional noise introduced by the optomechanical damping mechanism.  Note however that it is the
\emph{asymmetry} in the noise which leads to cooling or heating. The noise of the radiation pressure force $\hat F = \hbar
A\hat n$ 
 at positive frequency, $S_{FF}(+\omegam)$,
corresponds to the ability of the cavity to absorb a quantum of energy from the cantilever, while noise at
negative frequency $S_{FF}(-\omegam)$ corresponds to the ability of the cavity to emit a quantum of energy into
the cantilever.  Assuming $\omegam\gg\Gamma_M$, the net optical damping rate of the cantilever from Fermi's
Golden Rule is
\be
\Gamma_{\rm opt} = \frac{1}{\hbar^2} \left[S_{FF}(\omegam) - S_{FF}(-\omegam)\right] x_{\ZPF}^2 \label{Gammaopt}
\ee
where $x_{\ZPF}$ is the cantilever zero point position uncertainty, and $S_{FF}=\hbar^2 A^2 S_{nn}$.

As can be seen from Eq.~(\ref{eq:Snnomega}) and in Fig.~(\ref{fig:CavityTeff}a), for positive detuning, the noise
peaks at {\em negative} $\omega$ meaning that the noise tends to heat the degree of freedom $\hat x$.  For
negative detuning the noise peaks at positive $\omega$ corresponding to the cavity absorbing energy from $\hat
x$. Essentially, the interaction with $\hat x$ (three wave mixing) tries to Raman scatter the drive photons into
the high density of states at the cavity frequency. If this is uphill in energy, then $\hat x$ is cooled. 
A somewhat similar mechanism involving an oscillator with non-linear damping was considered by Dykman \cite{1978_08_Dykman_HeatingCoolingOscillator}.

\begin{figure}[bht]
\begin{center}
\includegraphics[width=3.45in]{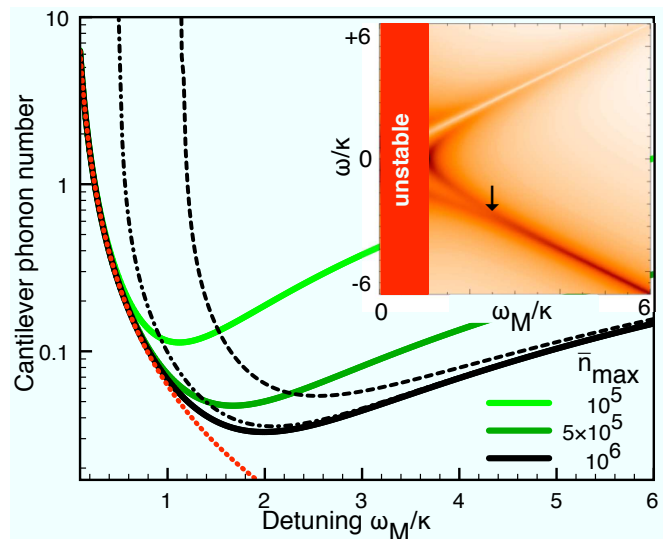}
\caption{(Color online) The steady state phonon number ${\bar n}_M$ obtained through cavity sideband cooling as a function of
detuning, with $\Delta=-\omega_M$. Thick lines represent the weak-coupling quantum noise result, Eq. (\ref{meanquanta}), for three
different laser powers (expressed via constant ${\bar n}_{\rm max}$, so the actual photon number $\bar n$ increases towards smaller $\omega_M$). The
bath temperature corresponds to ${\bar n}^T_M=100$ phonons, and the optomechanical coupling has been fixed to yield
a realistic ratio $\Gamma_{\rm opt}({\bar n}=1,\omega_M\rightarrow\infty)/\Gamma_M=0.1$. The dotted red line shows  ${\bar
n}^O_M=(\kappa/(4\omega_M))^2$ from Eq. (\ref{minimumPhononNumber}). The dashed (dash-dotted) lines show the results of the input-output
equations, for $\Gamma_M/\kappa=10^{-4} (10^{-5})$ and ${\bar n}_{\rm max}=10^6$. They were obtained by integrating the
exact spectrum $S_{cc}(\omega)$, Eq. (\ref{spectrum}), which is shown in the inset  as a density plot, for $\Gamma_M/\kappa=10^{-4}$
(Arrow indicates splitting of resonance; ``unstable'' regime does not permit $\Delta=-\omega_M$). 
The weak-coupling limit is recovered for $\Gamma_M/\kappa \rightarrow 0$. }
\label{fig:FinalPhononNumber}
\end{center}
\end{figure}

The radiation pressure noise is completely non-equilibrium but can be assigned a unique effective temperature
assuming that the mechanical oscillator is perfectly harmonic with sufficiently weak optical and mechanical damping, $\Gamma_M + \Gamma_{\rm opt} \ll \kappa$, so that it responds to the noise only at frequencies
$\pm\omegam$.  In the absence of mechanical damping, the mean number of mechanical quanta $\nmO$ present in the
steady state of the cantilever is given by the detailed balance expression (for $\Delta < 0$)
\be
\frac{\nmO +1}{\nmO} = \frac{S_{FF}(+\omegam)}{S_{FF}(-\omegam)}=\exp\left({\hbar \omegam \over T_{\rm
eff}}\right).
\label{nmO}
\ee
The noise temperature $T_{\rm eff}$ will
become negative when the noise leads to heating (see Fig.~\ref{fig:CavityTeff}). From Eq.~(\ref{eq:Snnomega}) we
obtain
\be
\nmO = -\frac{(\omegam+\Delta)^2+(\kappa/2)^2}{4\omegam\Delta}.
\ee
For the special case of detuning $\Delta=-\omegam$, we have the simple limit
\be
 \nmO =
\left(\frac{\kappa}{4\omegam}\right)^2 \label{minimumPhononNumber}
\ee
which shows that in the limit $\omegam \gg
\kappa$, the quantum ground state can be approached provided that the optical damping dominates over the
mechanical damping.  For the case $\Delta=-\omegam$, we obtain from Eq. (\ref{Gammaopt}):
\be
\Gamma_{\rm opt} = 4\left({x_{\ZPF} \over L}\right)^2\frac{\omega_R^2\bar n}{\kappa}
\frac{1}{1+\big(\frac{\kappa}{4\omegam}\big)^2}
\ee
If the mechanical damping $\Gamma_{\rm M}$ is not negligible compared to $\Gamma_{\rm opt}$, then a rate
equation yields the full expression for the mean number of mechanical quanta in steady state (Fig. \ref{fig:FinalPhononNumber}),
\be
\nm = \frac{\Gamma_{\rm opt} \nmO + \Gamma_{\rm M}\nmT}{\Gamma_{\rm opt} + \Gamma_{\rm M}} \label{meanquanta}
\ee
where $\nmT$ is the equilibrium mechanical mode occupation number determined by the mechanical bath temperature.
 In the limit of large mode occupation we can replace the
occupation numbers by the corresponding temperatures and we see that this expression reduces to the expected
classical expression for the final effective temperature.

We emphasize once more that using a large value of the detuning $|\Delta|=\omega_M\gg\kappa$ offers the
advantage of being able to reach an arbitrarily small minimum phonon number by optomechanical cooling, see Eq.
(\ref{minimumPhononNumber}). The only price to pay is to increase the  input intensity, in order to keep
constant the number of photons inside the cavity,
$
\bar{n}={\bar{n}_{\rm max}}/(1+(2\Delta/\kappa)^{2}),
$
which determines the cooling rate.
Here $\bar{n}_{{\rm max}}$ is the photon number at resonance, proportional to the input power. (Note however that $\bar{n}$ 
is limited by our weak coupling assumption $\Gamma_{\rm opt} + \Gamma_M \ll \kappa$)  In the good
cavity limit, the optimum cooling condition $\Delta=\omega_M$ requires detuning the drive laser by more than a
cavity linewidth. Although this decreases the cooling efficiency in terms of the input power, it
\emph{increases} the efficiency in terms of the circulating power. This is advantageous in experiments where the
limiting factor is heating of the cantilever by residual absorption from the circulating power. 



Apart from the cooling or heating of the cantilever, there is the well-known optomechanical frequency shift of
the mechanical mode (``optical spring effect'') (see e.g.
\cite{2002_02_BraginskyVyatchanin_Tranquilizer,2006_12_NergisMavalvala_LIGO}), which can also be expressed in
terms of the force spectrum, using second-order perturbation theory:
\begin{equation}
\delta\omega_{M}=\frac{x_{\ZPF}^{2}}{\hbar^{2}}\int\frac{d\omega}{2\pi}S_{FF}(\omega)
\left[\frac{1}{\omega_{M}-\omega}-\frac{1}{\omega_{M}+\omega}\right].\label{freqshift}
\end{equation}
We emphasize that Eqs. (\ref{Gammaopt}), (\ref{nmO}), (\ref{meanquanta}), and (\ref{freqshift}) can be used to
obtain the optical spring effect, the cooling rate, and the cooling limited phonon number for arbitrary force
noise spectra. Such more complicated spectra might, for example, result from the contribution of more than one
mode, or relate to a model different from the one discussed here, e.g. the coupling of the cantilever to some
electrical circuit \cite{2006_06_Wineland_CantiCoolingRFCircuit}. A related discussion in the context of cooling
by a single-electron transistor has been provided in
Ref.~\cite{2005_11_Clerk_SSET_Cooling}.

We now compare with commonly employed simpler models \cite{2004_12_ConstanzeKhaled_WithNote,
2006_06_Wineland_CantiCoolingRFCircuit, 2006_05_AspelmeyerZeilinger_SelfCoolingMirror}, where one postulates the cavity light intensity to approach its position-dependent equilibrium
value in an exponential fashion, 
$
{dn / dt} = \gamma ({\bar n}(x) - n), \label{relaxeq}
$
with ${\bar n}(x)$ denoting the equilibrium photon number as a function of position. 
Linearizing this equation and solving for the motion of $x$ yields an effective optomechanical damping
rate \cite{2004_12_ConstanzeKhaled_WithNote, 2006_06_Wineland_CantiCoolingRFCircuit,
2006_05_AspelmeyerZeilinger_SelfCoolingMirror}
\begin{equation}
\Gamma'_{\rm opt}={\hbar A^2 \over m \gamma} {1 \over 1+(\omega_M/\gamma)^2 } {\partial {\bar n} \over \partial
\Delta}. \label{simplemodelGamma}
\end{equation}
In general, this is {\em not} equal to the correct quantum-mechanical result (\ref{Gammaopt}), if we use a fixed
$\gamma\propto \kappa$, and it would predict an optimum damping rate for $\gamma$ comparable to $\omega_M$.  For
damping by radiation pressure, the full analysis shows that we must instead employ a detuning-dependent
effective decay rate, $\gamma\equiv (({\kappa \over 2})^2 + \Delta^2)/\kappa$. Then we recover Eq.
(\ref{Gammaopt}) in the limit $\omega_M \ll \kappa$. We note that Eq. (\ref{simplemodelGamma}) is always valid
when applied to bolometric forces, where one describes the time-lag of the
bolometric force due to a finite heat relaxation rate $\gamma\ll\kappa$, and where a quantum theory would be
different from the one presented here due to the dissipative nature of the force.

The quantum result (\ref{Gammaopt}) for $\Gamma_{\rm opt}$ can be reproduced by a classical theory based on
linearizing the equation of motion for the complex light amplitude $a$ itself
\cite{2002_02_BraginskyVyatchanin_Tranquilizer,2006_11_Kippenberg_RadPressureCooling}, of the form
$ {d a / dt} = i(\Delta+A x) a - {\kappa \over 2} (a - {\bar a}).
$
Of course it is still not possible to recover the correct formula (\ref{meanquanta}) for the
steady-state phonon number within such a purely classical theory, unless it is extended to include the
zero-point fluctuations.


In order to have direct access to the mechanical and optical fluctuation spectra, we now derive an exact
solution of the linearized Heisenberg equations of motion for $\hat{d}$ and $\hat{c}$, where we use the
input-output formalism \cite{1995_Walls_Milburn_QuantumOpticsBook} to take into account the damping and the
fluctuations, both for the optical and the mechanical degree of freedom (see also
\cite{2006_06_Paternostro_NJP_Spectrum}):
\begin{eqnarray}
\dot{\hat{d}} & = & i\Delta\hat{d}-\frac{\kappa}{2}\hat{d}-\sqrt{\kappa}\hat{d}_{{\rm in}}+i\alpha(\hat{c}+\hat{c}^{\dagger})\\
\dot{\hat{c}} & = & -i\omega_{M}\hat{c}-\frac{\Gamma_{M}}{2}\hat{c}-\sqrt{\Gamma_{M}}\hat{c}_{{\rm
in}}+i(\alpha^{*}\hat{d}+\alpha\hat{d}^{\dagger}).\end{eqnarray} Here the effective light amplitude
has been expressed in terms of a frequency, $\alp={\bar a}(\omega_{R}x_{\ZPF}/L)$ with $|{\bar a}|^2={\bar n}$. The solution yields the cantilever spectrum $S_{cc}(\omega)=\int dt\, e^{i\omega t}\left\langle \hat{c}^{\dagger}(t)\hat{c}\right\rangle $:

\begin{equation}
S_{cc}(\omega)=\frac{\Gamma_{{\rm M}}\sigma_{{\rm th}}(\omega)+\frac{|\alpha|^{2}}{\kappa}\sigma_{{\rm opt}}(\omega)}{|\mathcal{N}(\omega)|^{2}}, \label{spectrum}\end{equation}
where 

\begin{eqnarray}
\sigma_{{\rm th}}(\omega) & = & (\bar{n}_{{\rm M}}^{{\rm T}}+1)|\Sigma(\omega)|^{2}+\bar{n}_{{\rm M}}^{{\rm T}}|\chi_{M}^{-1}(\omega)+i\Sigma(\omega)|^{2}\\
\sigma_{{\rm opt}}(\omega) & = & \kappa^{2}|\chi_{R}(\omega)|^{2}|\chi_{M}^{-1}(\omega)|^{2}\\
\mathcal{N}(\omega) & = & \chi_{M}^{-1}(\omega)\chi_{M}^{-1*}(-\omega)+2\omega_{{\rm M}}\Sigma(\omega).\end{eqnarray}
We introduced the response functions of mirror and optical resonator, 
$\chi_{M}(\omega)=[\frac{\Gamma_{M}}{2}-i(\omega-\omega_{M})]^{-1}$ and
$\chi_{R}(\omega)=[\frac{\kappa}{2}-i(\omega+\Delta)]^{-1}$, and we defined the optomechanical {}``self-energy''
$\Sigma(\omega)=-i|\alp|^{2}(\chi_{R}(\omega)-\chi_{R}^{*}(-\omega))$.

The quantum noise results given previously are valid in the weak-coupling limit $\Gamma_M,\Gamma_{\rm opt}\ll\kappa$. Then, the optomechanical 
damping and the {}``optical spring'' frequency shift can be read off the self-energy as ${\rm
Im}\Sigma(\omega_{M})=-\Gamma_{{\rm opt}}/2$ and ${\rm Re}\Sigma(\omega_{M})=\delta\omega_{M}$, coinciding with
the expressions given above (Eqs. (\ref{Gammaopt}) and (\ref{freqshift})). The steady state
average phonon number $\bar{n}_{{\rm M}}=\int\frac{d\omega}{2\pi}S_{cc}(\omega)$ reproduces Eq.
(\ref{meanquanta}). The optical output spectrum (of $\hat{d}_{{\rm out}}=\hat{d}_{{\rm
in}}+\sqrt{\kappa}\hat{d}$) displays a Stokes peak at $\omega=-\omega_{M}$ and an anti-Stokes peak at
$\omega=+\omega_{M}$, with weights given by $\Gamma_{{\rm opt}}(\bar{n}_{M}+1)\bar{n}_{M}^{O}$ and $\Gamma_{{\rm
opt}}\bar{n}_{M}(\bar{n}_{M}^{O}+1)$, respectively, which are the rates of processes leading to heating/cooling
of the cantilever by red-/blue-shifting a reflected photon.

The exact solution allows us also to discuss the regime of strong cooling.
In the limit $\Gamma_{{\rm opt}}\approx4|\alpha|^{2}/\kappa\gg\Gamma_{{\rm M}}$, 
it adds a term $\bar{n}_{{\rm M}}^{{\rm T}}\Gamma_{{\rm M}}/\kappa+2\bar{n}_{{\rm M}}^{{\rm O}}\Gamma_{{\rm opt}}/\kappa$
to Eq. (\ref{meanquanta}) for $\bar{n}_{{\rm M}}$ (assuming $\kappa,\alpha\ll\omega_{M}$
and $\Gamma_{{\rm M}}\ll\kappa$). Then a \emph{minimal} phonon number
$\bar{n}_{{\rm M}}^{{\rm min}}\geq\bar{n}_{{\rm M}}^{{\rm T}}\Gamma_{{\rm M}}/\kappa$
is found as a function of $\Gamma_{\rm opt}$ at $(\Gamma_{{\rm opt}}/\kappa)^{2}=(\Gamma_{{\rm M}}/\kappa)(\bar{n}_{{\rm M}}^{{\rm T}}/\bar{n}_{{\rm M}}^{{\rm O}})/2$.
For $\Gamma_{{\rm opt}}/\kappa>1/2$, the mirror resonance peak splits
into a pair of peaks at $-\omega_{{\rm M}}\pm\alpha$, getting hybridized
with the driven cavity mode (see arrow in inset of Fig. \ref{fig:FinalPhononNumber}). At even
larger photon number or smaller $\omega_M$, when $\alpha\approx\omega_M/2$, the static bistability  \cite{1983_10_DorselWalther_BistabilityMirror} precludes
reaching the desired detuning $\Delta=-\omega_M$ (see inset of Fig. \ref{fig:FinalPhononNumber}).
Thus, we note that the far-detuned regime $\omega_M\gg\kappa$ has the additional strong advantage of avoiding the bistability,
which already interferes with cooling in some current schemes \cite{2006_11_Kippenberg_RadPressureCooling}. 

We have obtained the full quantum theory of cavity assisted sideband cooling of a cantilever, based on the
quantum noise approach applied to the fluctuations of the radiation pressure. In the previously unexplored
regime of a detuning much larger than the cavity linewidth, the cantilever may be cooled to arbitrarily small
phonon numbers. The theory analyzed here may form the basis for the interpretation of future optomechanical
experiments that will venture into the quantum regime of mechanical motion.

We thank M. Dykman and T. Kippenberg for comments, and particularly J. Harris for useful contributions.
This work was supported in part by the NSF under grants ITR-0325580, DMR-0342157, and DMR-0603369, by the NSERC, and by the SFB 631 of the DFG.

\bibliographystyle{apsrev}
\bibliography{/home/florian/pre/bib/shortall}

\end{document}